\documentclass[aps,prl,twocolumn,showpacs,preprintnumbers,amsmath,amssymb,superscriptaddress]{revtex4}

\usepackage{graphicx}
\usepackage{dcolumn}
\usepackage{bm}

\newcommand{\ty}[1]{\mbox{\tiny #1}}

\begin{document}

\title{Room-Temperature Superfluidity in Graphene Bilayers}

\author{Hongki Min$^*$, Rafi Bistritzer$^*$, Jung-Jung Su$^*$ and A.H. MacDonald}
\affiliation{Department of Physics, The University of Texas at Austin, Austin Texas 78712\\
\rm{(}$^*$These authors contributed equally to this work.\rm{)}}

\date{\today}

\begin{abstract}
Because graphene is an atomically two-dimensional gapless semiconductor
with nearly identical conduction and valence bands,
graphene-based bilayers are attractive candidates for high-temperature
electron-hole pair condensation.  We present estimates which suggest that the
Kosterlitz-Thouless temperatures of these two-dimensional
counterflow superfluids can approach room temperature.
\end{abstract}


\maketitle

\noindent
{\em Introduction}--- Electron-hole pair (exciton) condensates were first proposed\cite{Blatt1962,Keldysh}
as possible ordered states of solids more than forty years ago but have proved difficult to realize
experimentally.  Progress has been made recently
with the discovery\cite{Spielman2000,jpeahmnature} of equilibrium exciton condensation
below $T \sim 1 {\rm K}$ in the quantum Hall regime, the identification\cite{coldopticalexciton}
of spontaneous coherence effects in cold optically excited exciton gases, and studies of
dynamic condensation\cite{polaritonBEC} of polaritons in non-resonantly pumped optical microcavities.
In the weak-coupling limit
exciton condensation is a consequence of the Cooper instability\cite{Keldysh} of solids with
occupied conduction band states and empty valence band states inside identical Fermi surfaces.
Bilayer exciton condensates are counterflow superfluids with unusual
electrical properties\cite{quantumwell,Cavendish,jpeahmnature,Moon,avb,jungjung} which have so far been studied
experimentally mainly in the quantum Hall regime.
In this Letter we point out that superfluidity is likely to persist to remarkably high temperatures
in graphene based bilayers.
Graphene is a particularly attractive candidate for room
temperature bilayer exciton condensation because it is atomically two-dimensional, because it is a gapless
semiconductor, and because its two-dimensional massless Dirac band structure implies nearly
perfect particle-hole symmetry and stiff phase order.

\begin{figure}[h]
\includegraphics[width=0.9\linewidth]{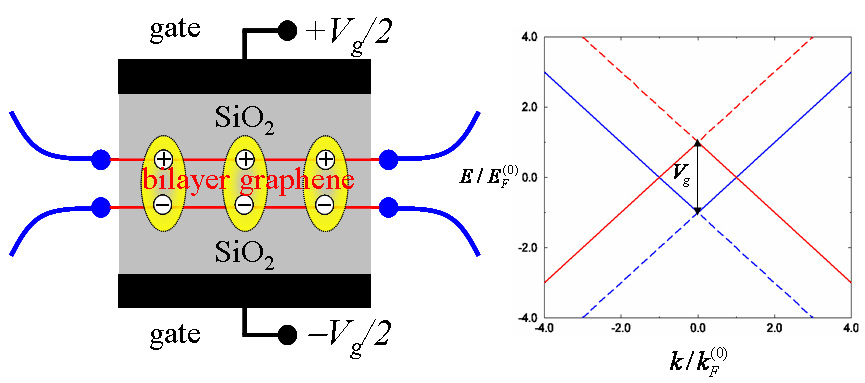}
\caption{(Color online) Left: Schematic illustration of a graphene bilayer exciton condensate channel in which
two single-layer graphene sheets are separated by a dielectric ($\rm{SiO_2}$ in this illustration) barrier.
We predict that electron and hole carriers induced by external gates will
form a high-temperature exciton condensate. Right: The two band model in which the two remote bands
indicated by dashed lines are neglected.}
\label{fig:system}
\end{figure}

We consider a system with two graphene layers embedded in a
dielectric media and gated above and below as illustrated schematically in Fig.~\ref{fig:system}.
Each layer has two Dirac-cone bands centered at inequivalent points in its Brillouin-zone.
The top and bottom gates can be used to control the electric fields $E_{ext}$ both above and below the bilayer.
When the two fields are equal the bilayer is neutral, but charge is transferred from one layer to the
other.  The Fermi level lies in the
graphene conduction band of one layer (the n-type layer) and in the valence band of the other layer (the p-type layer).
The particle-hole symmetry of the Dirac equation ensures perfect nesting\cite{nestingcaveat}
between the electron Fermi spheres in the n-type layer and its hole counterparts in the opposite layer,
thereby driving the Cooper instability.    The condensed state establishes spontaneous long-range coherence
between the two graphene layers.

Our main interest here is in providing an estimate of the maximum
possible Kosterlitz-Thouless (KT) temperature $T_{\ty{KT}}$
of these two-dimensional counterflow superfluids\cite{jpeahmnature}.
We use a two band model\cite{twobandcaveat} in which the occupied valence band of the n-type layer and the empty conduction
band of the p-type layer are neglected.
Our $T_{\ty{KT}}$ estimate is constructed from mean-field (Hartree-Fock) theory calculations\cite{nobeclimit} of the
temperature dependent phase stiffness of the ordered state.

\begin{figure}[h]
\includegraphics[width=0.9\linewidth]{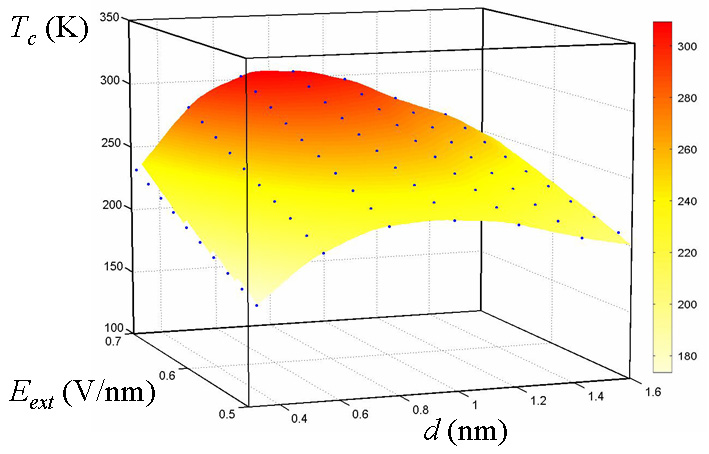}
\caption{(Color online) Normal to superfluid phase diagram showing the dependence of the critical temperature $T_c$
in Kelvin on the
distance between layers $d$ in nm and external bias electric field $E_{ext}$ in V/nm.}
\label{fig:phase3D_KT}
\end{figure}

Our main result is the normal to superfluid phase boundary depicted in Fig.~\ref{fig:phase3D_KT}.
The KT temperature is plotted as a function of the separation between the layers $d$ and
the electric field $E_{ext}$ outside the bilayer.
We estimate that superfluidity can survive at room temperature under favorable experimental conditions.
The non-monotonic dependence of $T_{\ty{KT}}$ on
$d$ at fixed $E_{ext}$ follows from a competition between the increasing
carrier density and the decreasing strength
of interlayer electron-hole interactions with increasing $d$.
At small $d$ the phase stiffness is limited by the carrier concentration,
which increases with $d$.  At large $d$, the KT temperature is limited by the same fermion-entropy
effects which are responsible for the Bardeen-Cooper-Schrieffer (BCS)
critical temperature of weak-coupling superconductors.

\noindent
{\em  Two-band mean-field theory}---
In the band eigenstate representation our band-Hamiltonian is
${\cal H}^{\ty{B}}= -\sum_{{\bm k},\sigma',\sigma}  c^{\dagger}_{{\bm k},\sigma'} \epsilon_k \tau^z_{\sigma' \sigma} c_{{\bm k},\sigma}$,
where $\epsilon_k = V_g/2-\hbar v  k$, $v$ is the band quasiparticle velocity,
$\bm{\tau}$ is a Pauli matrix vector which acts on the {\em which layer} pseudospin,
and $V_g = eE_{ext}d$ is the gate induced potential difference between the two layers.

Spontaneous interlayer coherence is induced by interlayer Coulomb interactions.
In the mean-field description the interlayer interaction reorganizes
the low-energy fermionic degrees of freedom into quasiparticles which are phase coherent
linear-combinations of the single-layer states.
The mean-field theory Hamiltonian can be written in the following form \cite{min2008}:
\begin{equation}
{\cal H}^{\ty{MF}} = -\sum_{{\bm k},\sigma',\sigma}  c^{\dagger}_{{\bm k},\sigma'}
\big[ \Delta^0_{\bm k} \delta_{\sigma',\sigma}+ {\bm \Delta}_{\bm k} \cdot {\bm \tau}_{\sigma',\sigma} \big]
c_{{\bm k},\sigma}.
\label{eq:HF}
\end{equation}
Because of the model's particle-hole symmetry $\Delta^0$ vanishes.
The pseudospin effective field ${\bm \Delta}_{\bm k}$ in Eq.~(\ref{eq:HF}) solves the following
self-consistent equation:
\begin{eqnarray}
\Delta^z_{\bm k}&=&\epsilon_k
+\frac{1}{2A} \sum_{\bm{p}}\left[ V^{(\ty{S})}_{\bm{k},\bm{p}} - \frac{2\pi e^2}{\epsilon} g d \right]  \left[ 1+{f_d(\Delta_{\bm p})} n_z(\bm{\Delta}_{\bm{p}}) \right]
\nonumber \\
\bm{\Delta}^\perp_{\bm k} &=& \frac{1}{2A} \sum_{\bm{p}} V^{(\ty{D})}_{\bm{k},\bm{p}} \; f_d ({\Delta_{\bm{p}} })  \bm{n}^\perp( \bm{\Delta}_{\bm{p}})
\label{eq:B}
\end{eqnarray}
where $A$ is the area of a graphene layer,
$\bm{\Delta}^\perp_{\bm k}=(\Delta^x_{\bm k},\Delta^y_{\bm k})$,
${\bm n}$ is a unit vector parallel to $\bm{\Delta_k}$, $g=4$ accounts for
the spin and valley degeneracy, and $f_d(x) = \tanh(x/2T)$ is the difference
between the occupation numbers of the negative energy and positive energy quasiparticles.
The Coulomb matrix element of the intralayer  interactions in the eigenstate basis is
\begin{equation}
V^{(\ty{S})}_{\bm{k},\bm{p}}  = \frac{1}{\epsilon} {2\pi e^2 \over |{\bm k}-{\bm p}|} \,{1+\cos(\phi_{{\bm k}}-\phi_{\bm p}) \over 2}      \label{eq:Vs}
\end{equation}
where  $\epsilon$ is the dielectric constant characterizing the embedding media,
and $\phi_{\bm k}=\tan^{-1}(k_y/k_x)$.
The corresponding matrix element of the interlayer interaction is
$V^{(\ty{D})}_{\bm{k},\bm{p}}=V^{(\ty{S})}_{\bm{k},\bm{p}} \exp(-|\bm{k-p}|d)$.
All energies are measured relative to the Dirac-point
chemical potential of the balanced bilayer\cite{explain1}.
Note that each spin and valley pairs independently and that
electron-hole condensation is indifferent to
spin-valley space rotation in either layer.

The interaction strength in a graphene monolayer is usually
characterized by the dimensionless effective fine structure constant,
$\alpha=
e^2/\epsilon \hbar v$.
This constant naturally appears in Eq.~(\ref{eq:B}) if energies and momentum are expressed
in units of $\hbar v k_{\ty F}^{\ty{(0)}}$ and $k_{\ty F}^{\ty{(0)}}$ respectively.  Here
$\hbar v k_{\ty F}^{\ty{(0)}} = V_g/2$ is the band Hamiltonian Fermi momentum.
The strength of the interlayer interaction is determined by $\alpha$ and by $k_{\ty F}^{\ty{(0)}} d$.

\begin{figure}[h]
\includegraphics[width=0.8\linewidth]{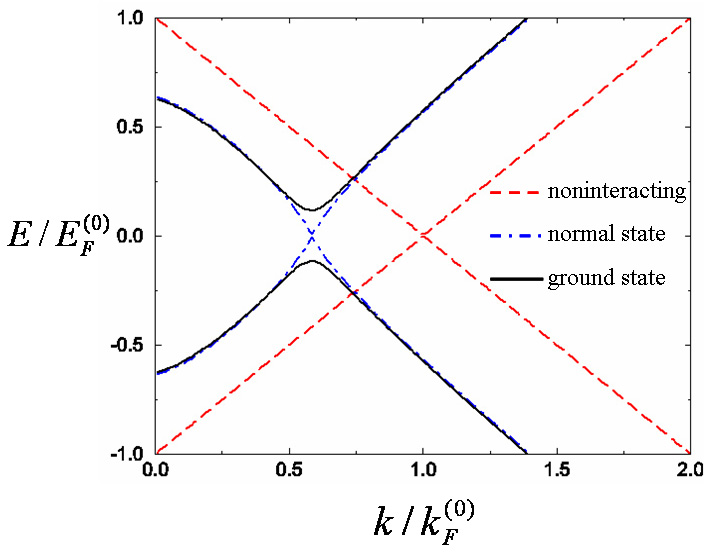}
\caption{(Color online) Mean-field theory energy bands for $\alpha=1$, $T=0$ and $k_{\ty{F}}^{(0)}d=1$.
($E_{\ty{F}}^{(0)}=\hbar v k_{\ty{F}}^{(0)}=V_g/2$.)
Note that $k_{\ty{F}} < k_{\ty{F}}^{(0)}$ because $E_{ext}$ is screened.}
\label{fig:projection}
\end{figure}

Interestingly, the self-consistent equations (\ref{eq:B}) admit solutions with non-zero chirality $J$ of the gap function $\bm{\Delta}^\perp$:
$\bm{\Delta}^\perp_{\bm k} = \Delta^\perp_k\left( \cos(J \phi_{\bm k}), \sin(J \phi_{\bm k}) \right)$. However, the
critical temperature of a state with non-zero chirality is higher than that of the corresponding $T_c$ of the zero chirality ground state
so these solutions are unlikely to be physically relevant. We focus on the $J=0$ solutions hereafter.

In the normal state, there is no interlayer coherence so $\bm{\Delta}^\perp$ vanishes. The intralayer Hartree-Fock potential
then follows from self-consistent solution for $\Delta^z$.  The main effects of electron-electron interactions in this case are to increase the
bare quasi-particle velocity\cite{yafisprl} and to screen the external bias voltage.  Screening reduces the amount of
charge transfer and therefore reduces the normal state Fermi momentum.
As illustrated in Fig.~\ref{fig:projection} the energy bands change qualitatively
in the condensed state because interlayer interactions induce coherence
between the two layers and open an energy gap.

\noindent
{\em  Linearized gap equation}---
The mean-field theory phase boundary between the normal phase and superfluid phase
is obtained by solving the linearized gap equation
\begin{equation}
\bm{n}^\perp(\bm{\Delta_k}) = \frac{1}{A}\sum_{\bm{p}} M_{\bm{k,p}} \  \bm{n}^\perp( \bm{\Delta_p})
\label{eq:gap_equation}
\end{equation}
obtained by linearizing Eq.~(\ref{eq:B}) with respect to $\bm{\Delta}^\perp$. The kernel
\begin{equation}
M_{\bm{k,p}} = \frac{1}{2 \Delta^z_{\bm k}} V^{(\ty{D})}_{\bm{k},\bm{p}} f_d (\Delta^z_{\bm p})
\label{eq:gap_equation_kernel}
\end{equation}
of the linearized gap equation is obtained by solving the self-consistent equation for $\Delta^z$
in the normal phase.
The normal phase is stable provided that all the eigenvalues of $M$
are smaller than one. By numerically evaluating $M$ for various interlayer distances and external fields
we find the mean-field phase diagram $T^{\mbox{\tiny MF}}_c(d,E_{ext})$ (not shown).

%


\noindent
{\em Phase stiffness}---
In two-dimensional superfluids the critical temperature is often substantially
overestimated by mean-field theory and is ultimately limited by
entropically driven vortex and antivortex proliferation at the
KT temperature
\begin{eqnarray}
T_{\ty{KT}} = \frac{\pi}{2}\rho_s(T_{\ty{KT}}).         \label{eq:T_KT}
\end{eqnarray}
We estimate $T_{\ty{KT}}$ by using mean-field theory to calculate
the phase stiffness (superfluid density) $\rho_s(T)$.
In parabolic band systems, this procedure yields reasonable estimates of $T_{\ty{KT}}$
in both BCS and BEC limits.

The phase stiffness is most easily calculated by evaluating the counterflow current
$\bm{j}_{\bm{Q}}=(e/\hbar)\rho_s \bm{Q}$ at small exciton momentum $\bm{Q}$.
Put formally, we evaluate the expectation value of the counterflow current operator
\begin{equation}
j^{\ty{D}}_{\bm{Q}} = -\frac{e v}{\small A} \sum_{\bm{k} \sigma} \cos\phi_{\bm{k}} \langle c^\dagger_{\bm{k},\sigma} c_{\bm{k},\sigma} \rangle  \label{eq:j_D}
\end{equation}
with the density matrix defined by the mean-field Hamiltonian
\begin{equation}
{\cal H}^{\ty{MF}} = {\cal H}^{\ty{B}} + \sum_{\bm{k}} \left( \Delta^{\perp}_{\bm{kQ}} c^\dagger_{\bm{k+Q/2},\uparrow} c_{\bm{k-Q/2},\downarrow} + h.c. \right),
\label{eq:Hamiltonian}
\end{equation}
where $\Delta^{\perp}_{\bm{kQ}}$  is the finite momenta pairing potential.

Placing $\bm{Q}$ along $\hat{\bm{x}}$ we find that
$\Delta^0_{\bm k} \to \frac{1}{2}\hbar Q v \cos\phi_k$ and that
\begin{equation}
j^{\ty{D}}_{\bm{Q}}=\frac{e v Q}{4\pi}\int dk \left[ \hbar v k \frac{\partial f(\Delta_{k})}
{\partial \Delta_{k}}- \frac{1}{2}f_d\left(\Delta_{k}\right) \hat{n}_z\left(\Delta_{k}\right) \right] \label{eq:j_Q_Delta}
\end{equation}
($\Delta_{k}^z=\epsilon_{k}$).
This expression for $j^{\ty{D}}_{\bm{Q}}$ has an ultraviolet divergence and fails to vanish in the normal state ($\Delta^{\perp} \to 0$).
Both properties are pathologies of the Dirac model.  When the two Fermi circles are
shifted in opposite directions at finite $Q$ they are asymmetric
with respect to the momentum-space origin.
As a consequence an ultraviolet cutoff at some momentum magnitude yields a finite counterflow current.
This current would vanish if the same calculation was performed
using a microscopic model with integrations over the full Brillouin-zone.  As long as $\Delta^{\perp}$ is small
compared to graphene's $\pi$-band width, a condition that is very easily satisfied, the anomalous ultraviolet contribution
to $\rho_s(T)$ is identical in the normal and in the condensed states.  It follows that the physical counterflow current
is related to the Dirac model counterflow current ($j^{\ty{D}}$) by
$j_{\bm{Q}} = j^{\ty{D}}_{\bm{Q}}(\Delta)-j^{\ty{D}}_{\bm{Q}}(\Delta^{\perp}=0)$.
Following this prescription, we conclude that the last term in Eq.~(\ref{eq:j_Q_Delta}) can be neglected and find that
\begin{eqnarray}
\rho_s(T) \approx \frac{v^2 \hbar^2 }{16\pi T} \int k dk \left[  {\rm sech}^2 \left( \frac{\epsilon_k}{2T} \right)
-{\rm sech}^2 \left( \frac{\Delta_k}{2T} \right)  \right].     \label{eq:rho_sT}
\end{eqnarray}
Note that the zero temperature phase stiffness,
\begin{equation}
\rho_s(T=0) \approx \frac{E_F}{4\pi},           \label{eq:rho_s_T0}
\end{equation}
is purely a normal state property just as in BCS theory.  Indeed an identical result is obtained
in the BCS theory of a parabolic band system when $\rho_s$
is expressed in terms of the Fermi energy.

An alternative approach for estimating $\rho_s(T)$ which also accounts for the intralayer interactions is to evaluate
the density matrix in Eq.~(\ref{eq:j_D}) using the self-consistent mean-field equations with finite pairing momentum.
As explained above, the physical counterflow current is obtained by subtracting $j_{\bm{Q}}^{\ty{D}}(\Delta^{\perp}=0)$ from $j_{\bm{Q}}^{\ty{D}}$.
The KT temperatures which follow from this procedure and Eq.~(\ref{eq:T_KT}) are depicted in Fig.~\ref{fig:phase3D_KT}.
Since $\rho_s(T)$ is a decreasing function of $d$ it follows from Eqs.(~\ref{eq:T_KT},\ref{eq:rho_s_T0}) that
$T_{\ty{KT}} \le E_{\ty{F}}/8$.  In our calculations we find that this inequality approaches
an equality when $k_{\ty{F}} d$ is small.  Consequently, the increase in $T_{\ty{KT}}$ with $d$
at small $d$ in Fig.~\ref{fig:phase3D_KT} simply follows the increase in $E_{\ty{F}} \sim e E_{ext} d/2$.

\noindent
{\em Discussion}--- The high transition temperatures we predict deserve comment. They are
larger than those of typical superconductors because condensation is driven by
Coulomb interactions over the full band width, rather than by phonon-mediated
interactions between quasiparticles in a narrow shell around the Fermi surface.
In this sense exciton condensation is more akin to ferromagnetism, which is also driven by
Coulomb interactions and can survive to very high temperatures.  The temperatures at which
exciton condensation can be achieved in graphene bilayers are immensely higher than those
which might be possible in semiconductor bilayers because more carriers can be induced
by external electric fields when the semiconductor has no gap, because the
Fermi energy increases more rapidly with carrier density for Dirac bands than
for parabolic bands, and because graphene
layers are atomically thin - eliminating the layer thickness effects which substantially
weaken Coulomb interaction in semiconductor quantum well bilayers.  The numerical
estimates reported in Fig.~\ref{fig:phase3D_KT} were obtained using a coupling constant
appropriate for a Si${\rm O}_2$ dielectric.  The optimal dielectric for high exciton
condensation temperatures should have a high dielectric breakdown field and a
low dielectric constant, suggesting that a suitable wide-gap material is likely the optimal choice.

Screening and other beyond-mean-field induced-interaction effects
are difficult to describe.  In the case of weakly interacting atomic gases
induced interaction effects can\cite{heiselberg} either increase or decrease
$T_c$, depending on the number of fermion flavors $g$.  For the present Coulomb interaction case,
a static Thomas-Fermi screening approximation with normal state screening wavevectors
reduces interaction strengths very substantially when spin and valley degeneracies ($g=4$) are included.
Mean-field-theory critical temperatures are reduced by a factor of $\sim e^{g}$
at small $d$ in this approximation and by a larger factor at large $d$.
On the other hand, when the screening wavevectors are evaluated in the condensed state
there is little influence on $T_{\ty{KT}}$ at small $k_Fd$
both because the large gap weakens screening and because
$T_{\ty{KT}}$ is proportional to the Fermi energy and not to the interaction strength in this limit.
All this leads us to suspect that at low-temperatures there is a first-order phase transition
as a function of layer separation $d$ between condensed and electron-hole plasma states,
similar to the transitions studied experimentally\cite{jpephasediagram} in quantum Hall exciton
condensates and theoretically\cite{senatore} in parabolic band bilayers.

Because of spin and valley degrees of freedom,
the exciton pairing we describe in this work is SU(4) symmetric; crudely speaking
the system has four identical superfluids simultaneously.  We therefore anticipate
interesting consequences of slightly unequal electron and hole densities,
similar to anticipated effects associated with the spin degree-of-freedom
in normal exciton condensates\cite{mypaperwithlenibascones,leonbalents}.
Because of this sensitivity, front and back gates which can control the electric fields
on opposite sides of the bilayer independently are highly desirable in
experimental searches for graphene bilayer exciton condensation.

Our finding that $T_{\ty{KT}} \sim 0.1 E_F$
in the limit of strong interactions between conduction band electrons and valence
band holes is partially supported by experimental studies\cite{shin} of fermionic cold atoms
in the strong-interaction unitary limit.  It implies that
$T_{\ty{KT}}$ should approach room temperature when $E_F$ is larger than $ \sim 0.3 {\rm eV}$
($n$ larger than $\sim 10^{13}{\rm cm}^{-2}$) and $d$ is smaller than $\sim 2 {\rm nm}$.
Experimental detection of spontaneous coherence through one of its characteristic
transport anomalies\cite{jpeahmnature} will be necessary to
construct a quantitatively reliable phase diagram.


\noindent
{\em Acknowledgment ---}This work has been supported by the Welch Foundation, by the Army Research Office,
by the NRI SWAN Center, and by the National Science Foundation under grant DMR-0606489.  AHM acknowledges
helpful discussions with Rembert Duine, Koos Gubbels, and Henk Stoof.



\begin{thebibliography}{999}
\bibitem{twobandcaveat}  In separate calculations not described here we found that
coherence between the two remote bands has little effect for $k_{\ty{F}} d > 1$ and
that will act to raise the KT temperature for $k_{\ty{F}} d < 1$.
\bibitem{nobeclimit}  In parabolic band systems this procedure provides a good estimate
of the critical temperature in both the weak-coupling BCS and the strong-coupling BEC limits.
Note however that because graphene is a gapless semiconductor, it does not have a simple BEC strong-coupling limit.
\bibitem{explain1} The unit contribution to the right-most square-bracket factor in Eq.~(\ref{eq:B}) for $\Delta_{\bm{k}}^z$
is due to exchange interactions with the full valence band of the n-type layer.
\end{thebibliography}
\end{document}